\documentstyle[epsfig]{mn}

\newcommand{\beq}{\begin{equation}}
\newcommand{\eeq}{\end{equation}}
\newcommand{\bey}{\begin{eqnarray}}
\newcommand{\eey}{\end{eqnarray}}

\newcommand{\kpc}{\, {\rm kpc} }

\newcommand{\grad}{{\bf \nabla}}

\newcommand{\be}{\begin{equation}}
\newcommand{\ee}{\end{equation}}
\newcommand{\ba}{\begin{eqnarray}}
\newcommand{\ea}{\end{eqnarray}}

\newcommand{\kms}{\, {\rm km}s^{-1}}
\newcommand{\r}{\mbox{\boldmath $r$}}

\newcommand{\thetab}{\mbox{\boldmath $\theta$}}

\newcommand{\nablab}{\mbox{\boldmath $\nabla$}}

\newcommand{\x}{\mbox{\boldmath $x$}}

\newcommand{\tg}{{g}}

\newcommand{\R}{\mbox{\boldmath $R$}}

\title[An Introduction to Gravitational Lensing in TeVeS] {
 An Introduction to Gravitational Lensing in TeVeS  }
\author[H. Zhao]{
    HongSheng Zhao$^{1,2}$, \\
 $^1$ SUPA, University of St. Andrews, KY16 9SS, UK \\
 $^2hz4@st-andrews.ac.uk$
} 
\date{Based on presentations at Sicily Lensing School, Oct 29-Nov 3.  
COMMENTS WELCOME}
\pagerange{\pageref{firstpage}--\pageref{lastpage}}
\pubyear{2006}
\begin{document}
\maketitle
\label{firstpage}
\begin{abstract}
Bekenstein's (2004) TeVeS theory has added an interesting twist to the search
for dark matter and dark energy, modifying the landscape of
gravity-related astronomy day by day.  Built bottom-up rather than
top-down as most gravity theories, TeVeS-like theories are healthily
rooted on empirical facts, hence immediately passing sanity checks on
galaxy rotation curves, solar system constraints, even bullet cluster
of galaxies and cosmology with the help of 2eV neutrinos.
Nonetheless, empirical checks are far from perfect and complete, and
groups of different expertises are rapidly increasing the number of
falsifiable properties of the theory.  The theory has also been made
much simpler and more general thanks to the work of Zlosnik, Ferreira,
Starkman (astro-ph/0606039, 0607411).  Here I attempt a tutorial of
how to compute lensing convergence, time delays etc 
in TeVeS-like theories for non-spherical lenses.  I gave examples  
to illustrate a few common caveats of Dark-Matter-guided intuitions.
\end{abstract}



\section{Introduction}

\subsection{GL in any co-variant metric theory}

While gravitational lensing (GL) is a cornerstone of Einsteinian
gravity, light propagation is actually well-defined in any generic
metric theories of gravity and one can test such theories using
lensing data.  In fact, light bending is a general property of E\&M
waves propagating following Fermat's principle, which happens in a
non-uniform medium where the effective speed of light $c$ varies
(e.g., as in atomospherical seeing) in a flat space-time.  Lensing
also happens as light with constant speed $c$ following geodicics in
vaccum in a curved space-time (e.g., bending by the Sun), but the link
between the matter density and the curvature need not be given
Einstein's equation.

For example, the gravitational pre-factor $G_{\rm eff}$ needs not be a
true constant in galaxies where the gravity is so weak that we lack
precise experiments to measure ${G_{\rm eff}m_1 m_2 \over r^2}$ force.
The solar gravity on Pluto $4 \times 10^4$ greater than a typical
place in a galaxy.  E.g., the Sun's acceleration around the Galaxy
\beq 
g \sim {(200 \kms)^2 \over 10 \kpc} 
  \sim {0.1 c \over HubbleTime}
  \sim {1 {\rm m} \over {\rm day}^2} 
  \sim {1 {\rm Angstrom} \over {\rm sec}^2} .
\eeq 

A mundane example of $1$Angstrom per second squared gravity is 
{\it the mutual Newtonian 
gravity of two nearly parallel sheets of printing papers} approximately.
It is well-known that a parallel plate electric capacitor yields a different 
E-field if immersed in vaccum entirely (Casmir effect), 
or filled in a dielectric air inside.  
The gravitational attraction of two sheets of paper could depend
on enviornment in a similar way for very different physics.
Consider a {\it Gedanken} experiment with 
a gravitationally torquing pendulum made by two 
misaligned suspended sheets of paper.  If one could measure
the period of the torquing pendumlum not only here on Earth 
(as in free-fall experiments in an Einstein tower), 
but also take the table-top experiments to 
the edge of the solar system (where Pioneer 10/11 probes are),
in the interstellar space (where galactic stars orbit)
and in the expanding void between galaxies, then 
one could measure how $G_{eff}$ changes with space and time.

\subsection{The three pillars of the standard cosmology}

The standard cosmological paradigm is built on three pillars: Cold
Dark Matter, a cosmological constant, and Einsteinian gravity.  While
independent experimental basis of each of the three is debatable on
astronomical scales, but their synergy (characterised by the
cosmological pie) has proven amazingly successful at describing the
Universe especially on large scale.

Despite its apparantly enticing simplicity, the paradigm has much to
be understood and is under pressure to be modified by observations of
galaxy scale and by competing theories like TeVeS.  For example, the
experimentally undetected cold dark matter (generally thought to be
SuperSymmetry particles) is predicted to clump in scale-free fashion,
while observations of dwarf galaxies suggest a kpc-scale
free-streaming length of dark matter particles.  The idea of
introducing a constant or a vaccum energy 100 orders of magnitude
lower than what the SUSY physics can provide naturally is still
regarded by many theoreticians as unsatisfactory.

These fine-tunings have lead some to believe the paradigm is an
effective theory, e.g., a 4D projection of a more fundamental 5D brane
world theory.  Some also question the Einsteinian gravity since its
associated equivalence principles, remain untested on galaxy scale and
cosmological scale.

\subsection{Modified gravity: motivations and history}

Modifying gravity is a reoccuring excercise which started ever since
the general acceptance of Einsteinian gravity, which was itself a
revolutionary modification to Newtonian gravity.  Many theories modify
the Einstein-Hilbert action to introduce a new scalar field which
manifests itself only through the extra bending of space time, but its
coupling to the metric is different from the simple coupling of
massive particles with the space-time metric.

By construction the theories would respect Special Relativity
prescription of metric co-variance, and preserve conservations of
momentum and energy.  They do allow for a table-top Cavendish-type
experiment with a torquing pendulum to measure an effective
gravitational constant $G_{eff}(t,x)$ which varies with time and
enviornment of the experiment.  For example, the recent motivation to
replace the cosmological constant in General Relativity leads to
theories with $G_{eff}$ depending on the curvature of space-time,
which evolves with the cosmic time in a way to drive the acceleration
of the universe at late time.

However, among two dozen theories proposed after GR, very few survive
the precise tests on SEP in the solar system and the well-studied
binary pulars.  Even fewer are motivated and succeeded in addressing
both astronomical dark matter and cosmological constant.

\subsection{The secret of TeVeS and variants}

TeVeS is an exception.  It holds 
the promise of explaining both dark matter and cosmological constant
by relaxing the SEP (strong equivalence principle) 
only in untested weak gravity envionments like in galaxies, but 
respecting the SEP to high accuracy in the solar system.

Crudely speaking such theory has an aether-like field with an
aquadratic kinetic term in its Lagrangian density, so the $G_{eff}$
can be made a function of the strength of gravity $|g|$, such that
$G_{eff}$ is constant within $10^{-16}$ anywhere in the solar system,
yet {\it varies} by a factor of 10 in galaxies and in the universe
over a Hubble time where $|g|$ is much smaller.  Enhancing the
$G_{eff}$ mimics the effects of adding dark matter, and reducing the
$G_{eff}$ can drive the acceleration of the universe.

\section{A characteristic scale for Dark Matter}

As one of the important issue to be understood about dark matter, it
has long been noted that on galaxy scales dark matter and baryonic
matter (stars plus gas) have a remarkable correlation, and respects a
mysterious acceleration scale $a_0$ (Milgrom 1983, McGaugh 2005).

The Newtonian gravity of the known matter (baryons etc.) 
${\bf g}_k$  and the dark matter gravity
${\bf g}_{DM}$ are correlated through an empirical relation
(Zhao and Famaey 2006, Angus, Famaey, Zhao 2006, 
Famaey, Gianfranco, Bruneton, Zhao 2006) 
such that the light-to-dark ratio, 
experimentally determined to fit rotation curves, satisfies
\begin{equation}
 {g_k \over g_{DM} } = { g_{DM} + \alpha g_k \over a_0 }, 
\end{equation}
where $0 \le \alpha \le 1$ is a parameter.  
Let $\alpha =0$ we get a very simple relation
\beq
g_{DM} \approx \sqrt{g_k a_0}, \qquad a_0 \equiv 1 Angstrom{\rm sec}^{-2}
\eeq
where $a_0$ is the forementioned gravity scale, below which DM and DE 
phenomena start to surface.

Such a tight correlation is difficult to understand in a galaxy
formation theory where dark matter and baryons interactions enjoy a
huge degrees of freedom.  This spiral galaxy based empirical relation
is also consistent with some elliptical galaxies and gravitational
lenses.

\section{A scale for dark energy}

Equally peculiar is
the amplitude of vaccum energy density $\Lambda$, which 
is of order $10^{120}$ times smaller than its natural scale.  
It is hard to explain from fundamental physics why vaccum energy
starts to dominate the Universe density only
at the present epoch, hence marking the present as the turning point for 
the universe from de-acceleration to acceleration.

This is related to the fact that 
\beq
a_0 \sim \sqrt{\Lambda} \sim c H_0
\eeq 
where $a_0$ is the characteristic scale of DM as well.  

{\it Somehow dark energy and dark matter are tuned to shift dominance when
the energy density falls below ${a_0^2 \over 8 \pi G}$}.  These
empirical facts should not be completely treated as random
coincidences of the fundamental parameters of the universe.
The explanation with standard paradigm has been unsatisfactory.

\section{The metric and dynamics of the TeVeS fields}

TeVeS, as GR, is a metric theory.  
Let $g_{\mu\nu}$ being the physical
coordinates, then near a quasi-static system like a galaxy, the physical space-time
is only slightly curved, and can be written as in a 
rectangular coordinate $\x=(x_1,x_2,x_3)$ centred on the galaxy as
 \ba
    -c^2 d\tau^2 &=& \tg_{tt} d{t}^2 + \tg_{rr} dl^2, \\
    dl^2&=&(dx_1^2+ dx_2^2+dx_3^2).
 \ea

Introducing a small quantity ${|\Phi| \over c^2} \ll 1$, we can
write the metric components
 \beq
    \tg_{rr} \approx  - c^2 \tg_{tt}^{-1}
    \approx  1 - {2\Phi \over c^2}.
 \eeq

To show that $\Phi(\x)$ takes the meaning of a gravitational
potential, we note that a non-relativistic massive particle
moving in this metric follows the geodesic
 \beq
    {d^2 x_i \over d{t}^2} - {\partial {g}_{tt} \over 2
    \partial x_i} \approx 0, \qquad
        \rightarrow \qquad {d^2 \x \over d{t}^2} \approx - \nablab
        \Phi(\x),
 \eeq
which is the equation of motion in the non-relativistic limit
where ${d{t} \over d\tau} \approx 1$. 

\subsection{Vector or scalar}

Near a quasi-static system like a galaxy, $g_{00} = -(1+ 2 \Phi)$, where we omit
the factor $c^2$ for clarity.
TeVeS predicts a time-like vector field with four components,
which are approximated as 
\beq
A^\alpha=(1 -\phi - \Phi, 0, 0 0)
\eeq
and
\beq
A_\alpha=-(1 -\phi + \Phi, 0, 0 0)
\eeq  
to the lowest order, where $\phi$ is a scalar field.

For most of the system that we are interested, the key
is the module of $A$, which is equivalently described by 
the scalar field $\phi$ related through
\beq
A^{2} \equiv g_{ab}A^{a}A^{b} \equiv -e^{-2\phi} < 0
\eeq

This shows the vector field is more fundamental than the scalar field
in TeVeS and TeVeS is can be described the physical metric and vector
field alone.
The original proposal of Bekenstein contains
two metric, while most recent work of Zlonik, Ferreira, Starkman 
(2006, PRD. 74, 0404037) shows that the theory is 
equally described by a single physical metric $g_{\mu\nu}$, 
whose geodicics particles and light will follow.  The other metric
(called Einstein metric) is fully 
described once the vector field is specified.

\subsection{TeVeS as dark matter}

In TeVeS, the galaxy potential $\Phi$ comes from two parts,
\beq
\Phi = \Phi_{kn} + \phi
\eeq
where the known Newtonian gravitational potential $\Phi_k({\bf x})$ 
of known matter of density $\rho_k({\bf x})$ satisfies
\beq
\grad \cdot \grad \Phi_{kn} = 4 \pi G \rho_{kn}
\eeq 
and the added scalar field satisfies
\beq
\grad \left[ \mu_s  \grad \phi \right] = 4 \pi G \rho_{kn},
\eeq
where
\beq
\mu_s= \left|{\grad\phi \over a_0} \right| 
+ O(\left|{\grad\phi \over a_0} \right|^2).
\eeq

The picture to keep in mind is that 
the scalar field replaces the usual role of the potential of the Dark Matter.
The vector field $A$ is fully specified once $\phi$ and $\Phi$ are given.

To illustrate how the scalar equation come from Lagrangian of 
the vector field theory, let's consider a toy model.

\section{An E\&M-like 4-vector potential}

Zlonik, Ferreira, Starkman (2006, astro-ph/0607411) 
generalised TeVeS as part of a broader class of Einstein-Aether theories,
which we will follow here.  We will neglect a certain Lagrangian 
multipler for normalisation of vector field 
and also make simplifications where possible 
(setting $c_1=c_3=-1$ and $c_2=0$ in their notation, and choosing
the simplest modification to the Langrangian).  
We emphasize the similarity of the 4-vector field here with the 4-vector potential 
$(A_0,A_1,A_2,A_3)$ in electromagnetism.

Let the vector field
$A_{\alpha} = g_{\alpha\beta}A^{\beta}$ be a time-like unit vector
with the constrain equation
\beq
A_{\alpha}A^{\alpha}=-1
\eeq
and let it be coupled to the metric $g_{\alpha\beta}$, 
so that the system is governed by an action $S$ or Lagrangian density $L$, 
\begin{eqnarray}
S &=& \int d^4x \sqrt{-g} 
\left[ { R \over 16 \pi G} + L_{kn} +  \left(1-{2f \over 3}+O(f^2)\right) L_f \right],  
\end{eqnarray}
where 
$R$ the usual Ricci scalar of the metric \textbf{g},
$L_{kn}$ is the known matter Lagrangian density coupling the known matter 
with metric but not coupled to the vector field directly.
The 3rd term is
the local Lagrangian density of the vector field, which 
apart from a dielectric-like modification factor $(1-2f/3+O(f^2))$, 
is the normal electromagnetism-like kinetic coupling to metric
\beq
L_f = {a_0^2 f^2 \over 32 \pi G } \equiv { F_{\alpha\beta}F^{\alpha\beta} \over 32 \pi G }
\eeq
where the dimensionless $f$ parameter 
is a measure of the strength of the Maxwell tensor field $F_{\alpha\beta}$: 
\beq
F_{\alpha\beta}=\partial_\alpha A_\beta - \partial_\beta A_\alpha.
\eeq
which is the covariant derivative of the 4-potential $A_\alpha$,
(as for the electric and magnetic field in electromagnetism). 

\subsection{Einstein eq. and vector field eq. of motion}

Taking variations of the action respect to the metric $g_{\alpha\beta}$ and $A_\alpha$
respectively we get the gravitational field equations and vector equation of motion
for this theory respectively.  Combining the two we have  
\bey
{G_{\alpha\beta} \over 8\pi G} &=& 
T^{kn}_{\alpha\beta} 
+ T_{\alpha\beta}, \\
\eey
where the left-hand side is the proportional to 
the Einstein tensor 
$G_{\mu\nu}  \equiv R_{\mu\nu}-{R \over 2} g_{\mu\nu}$ 
and on rhs the 1st term is the stress-energy tensor of known matter, 
the 1st term is the stress-energy tensor for the vector field 
${T}_{\alpha\beta}$, 
\bey
T_{\alpha\beta} &=& \tilde{T}_{\alpha\beta} 
+ {1 \over 8 \pi G} A_\alpha \nabla_\mu [ (f+O(f^2)) g^{\mu\nu} F_{\nu\beta} ],
\eey
which is a non-linear function of derivatives 
of the field $A_{\beta)}$.  
Note how {\it the vector field creates the effect of additional matter}.

Near a galaxy, $G_{00}= 2 \grad\grad \Phi$, so
the 00th moment of the above equation reduces to
\beq
\grad (\mu_s  \grad \Phi) = 4 \pi G \rho_{kn} ,
\eeq
where 
\beq
\mu_s = f+O(f^2), \qquad f= \left|{\grad \Phi \over a_0}\right|.
\eeq
This way we recover the classical MOND equation of 
Bekenstein and Milgrom (1984) in the {\it weak field limit} $f \rightarrow 0$,
i.e., 
the gravity $\grad \Phi$ drops as $\sqrt{GMa_0}/r$ far away from a point mass $M$.

\subsection{Hubble expansion equation}

For homogeneous flat cosmology, we can set the metric 
\beq
ds^2 = - c^2 dt^2 + a(t)^2  (dr^2+r^2 d\theta^2 + r^2\sin\theta^2 d\phi^2)
\eeq
and the vector field
\beq
A^\alpha = -A_\alpha = (1,0,0,0),
\eeq
and its derivatives
\beq
f=0.
\eeq
The Einstein equation reduces to the following eq. for Hubble expansion 
\beq
\left({da \over a dt}\right)^2  = {8\pi G \over 3} \rho_{kn}.
\eeq
Interestingly this is not modified from the GR, where $\rho$ is the density
of known matter of the uniform background.  

The actual theory of TeVeS and Zlonik et al. is more sophisticated
than above toy model.  E.g., 
Zhao \& Famaey (2006) proposed to modify TeVeS with a 
$\mu=f/(1+f)$, hence the Lagrangian
$\int (1-\mu) d{a_0^2 f^2 \over 32\pi G} = 
(f-\ln(1+f)){a_0^2 \over 16\pi G} = (1-{2f \over 3}+O(f^2) )L_f$, 
such that we recover Einstein-Newtonian in strong
gravity, and achieve a good fit to rotation curves in the intermediate
regime, and have a smooth transition between galaxies and cosmology.
The term $f^2$ or $L_f$ can contain derivatives other than the Maxwell tensors.
It is also possible to modify the homogeneous cosmology in
TeVeS-like theory to create the effect of dark energy.  The cosmology
prescribes distance-redshift relations, essential for lensing.  For
these we assume the background cosmology in TeVeS-like theories must
fit the same SNe distance data, hence the resulting distance-redshift
relation (at least at redshifts of order unity) is exactly as in LCDM.
A detailed study to this effect is shown in Zhao (2006,
astro-ph/0610925).

\section{The interpolating functions in MOND vs TeVeS}
 
In the classical MOND a spherical galaxy of 
rotation curves $V(r)$ could be explained by modifying gravity:
 \be
        g=\frac{V^2}{r}= \frac{GM}{r^2\mu(g/a_0)},
 \ee
where $\mu(g/a_0)$ is the effective ``dielectric constant", 
 depending on the gravitational field strength, $g$.  
The standard interpolating function 
\begin{equation}
        \mu(x)=\frac{x}{\sqrt{1+x^2}}
\end{equation}
is often used in fitting rotation curves.  But Zhao \& Famaey (2006)
argued that this function has undesirable features in TeVeS.  
Instead they (Angus, Famaey, Zhao 2006) proposed to use 
\begin{equation}
        \mu(x)=\frac{2x}{1+(2-\alpha x)+\sqrt{(1-\alpha x)^2+4x}}
\end{equation}
which is a parametric $\alpha$-family, which recovers
Bekenstein's (2004) toy model and the simple model of 
Famaey \& Binney (2005) by setting $\alpha=0$ and $\alpha=1$.

The gravitational potential in MOND theory satisfies a modified Poisson's  equation,
\begin{equation}
\nabla[ \mu \nabla \Phi] = \nabla^2 \Phi_N = 4 \pi G \rho_{kn}
\end{equation}
where the $\rho_{kn}$ is the density of all known matter.  This is 
different from TeVeS, where 
the total potential is the sum of Newtonian potential ($\Phi_N$) 
and a potential due to a scalar field ($\phi_s$):
\begin{equation}
\Phi=\Phi_N+\phi_s.
\end{equation}
We can see that the scalar potential plays the role of the dark matter
gravitational potential, and the Poisson-like equations for the
scalar field relates it to the Newtonian potential $\Phi_N$ (generated
by the baryonic matter),
\begin{equation}
\nabla[ \mu_s \nabla \phi_s] = \nabla^2 \Phi_N = 4 \pi G \rho
\end{equation}
where $\mu_s$ is a function of the scalar field strength $g_s=|\nabla \phi_s|$, and is derived from a free function in the action of the scalar field. 
In spherical symmetry, we have
\begin{equation}
\mu_s g_s=\mu(g_s+g_N)=g_N
\end{equation}
where $\mu$ is the interpolating function of MOND, hence the two 
interpolation function is related by 
\begin{equation}
\mu_s=\frac{\mu}{1-\mu}.
\end{equation}

\section{Light Bending in Slightly Curved Space Time}

Light rays trace the null geodesics of the space time metric.
Lensing, or the trajectories of light rays in general, are
uniquely specified once the metric is given.  In this sense light
bending works *exactly* the same way in any relativistic theory as
in GR.

Near a quasi-static system like a galaxy, the physical space-time
is only slightly curved.  Consider lensing by the galactic potential
$\Phi(\r)$.  A light ray moving with a constant
speed $c$ inside follows the null geodesics 
$d{t} = \sqrt{-\frac{\tg_{rr}}{\tg_{tt}}} dl$. 
An observed light ray travels a proper
distance $l_{os}=l_{ls}+l_{ol}$ from a source to the lens and then
to an observer. Hence it arrives after a time interval (seen by an
observer at rest with respect to the lens)
 \beq
       \int d{t} = \int_{0}^{l_{os}} {dl \over c} - \int_{0}^{l_{os}}
       {2\Phi({\bf r}) \over c^2} {dl \over c}
 \eeq
containing a geometric term and a Shapiro time
delay term due to the $\Phi$ potential of a galaxy. 

In fact, gravitational lensing in TeVeS recovers many familiar
results of Einstein gravity in (non-)spherical
geometries.  Especially an observer at redshift $z=0$ sees a
delay $\Delta t_{\rm obs}$
in the light arrival time due to a thin deflector at $z=z_l$
 \beq\label{tdr}
    {c\Delta t_{\rm obs}({\bf R}) \over (1+z_l)}  \approx
    {D_{s} \over 2D_{l}D_{ls}}
    \left({\bf R}- {\bf R}_s \right)^2
    - \int_{-\infty}^{\infty} \!\!\! dl {2\Phi({\bf R},l) \over c^2},
 \eeq
as in GR for a weak-field thin lens, $\Phi/ c^2 \ll 1$. A light
ray penetrates the lens with a nearly straight line segment (within the thickness of the lens) 
with the 2-D coordinate, $\R=D_l \thetab$, perpendicular to the sky, where $D_l(z_l)=l_{ol}/(1+z_l)$
is the angular diameter distance of the lens at redshift $z_l$,
$D_s$ is the angular distances to the source, and $D_{ls}$ is the
angular distance from the lens to the source.
The usual lens equation can be obtained from the gradient of the
arrival time surface with respect to $\R$.  i.e.,
\ba
x- {D_l D_{ls} \over D_s} \alpha_x(x,y) = x_s  \\\nonumber
y- {D_l D_{ls} \over D_s} \alpha_y(x,y) = y_s  
\ea
where the deflection 
\ba
\alpha_x & = & \int_{-\infty}^{\infty} \!\!\! dl {2 \partial_x \Phi(x,y,l) \over c^2},\\\nonumber
\alpha_y & = & \int_{-\infty}^{\infty} \!\!\! dl {2 \partial_y \Phi(x,y,l) \over c^2},
\ea
and the convergence
\beq
\kappa  = {D_l D_{ls} \over 2D_s }\left( \partial_x \alpha_x + \partial_y \alpha_y \right),
\eeq
and likewise for the shear,
\beq
\gamma_2 = D_l \partial_y \alpha_x , 
\eeq
and
\beq
\gamma_1 = {D_l \over 2}\left( \partial_x \alpha_x - \partial_y \alpha_y \right) 
\eeq
and for the amplification $A$
\beq
A^{-1} = (1-\kappa)^2 - \gamma_1^2 -\gamma_2^2.  
\eeq 
The time delay between a pair of images $i$ and $j$ 
is given by the path integral
\beq
{c \Delta t_{\rm obs}^{ij} \over (1+z_l)}  
= \int_{p_i}^{p_j} \!\!\! dp \, L(p),
\eeq
\bey
L(p) &\equiv&  {dx\over dp} 
\left({\alpha_x(p_i) + \alpha_x(p_j) \over 2} - \alpha_x(p)\right) \\\nonumber
&+& {dy\over dp} 
\left({\alpha_y(p_i) + \alpha_y(p_j) \over 2} - \alpha_y(p)\right),
\eey
where $x(p),y(p)$ defines a path from image $i$ to image $j$ 
as $p$ varies from $p_i$ to $p_j$.

\section{Subtle differences of TeVeS lensing and DM lensing}

\subsection{$\kappa$ of TeVeS and DM}

An interesting point is that in GR $\kappa$ is proprotional to the projected 
surface density of known matter.
This is {\it not the case for a non-linear theory}.  
We can express $\kappa$ into of the 
critical density as follows,
\beq
\kappa = {\tilde{\Sigma}(x,y) \over \Sigma_{crit}}, 
\qquad \Sigma_{crit}^{-1} \equiv {4 \pi G D_l D_{ls} \over D_s c^2 }, 
\eeq
where we define an effective projected density as follows,
\beq
\tilde{\Sigma}(x,y) \equiv  
\int_{-\infty}^{\infty} \!\!\! dl \tilde{\rho}(x,y,l),
\eeq 
note the integrand is NOT the true matter volume density at (x,y,l), 
rather
\beq
\tilde{\rho}(x,y,l) \equiv {\nabla^2 \Phi(x,y,l) \over 4 \pi G}
= \rho_{kn} +  \rho_{eDM} > \rho_{kn} 
\eeq
because $\Phi$ is addition of two fields, 
and we have an effective Dark Matter (eDM) from the $\phi$ field,
\beq
\rho_{eDM} ={\nabla^2 \phi(x,y,l) \over 4 \pi G} 
\eeq
The eDM tracks the known matter, because 
the TeVeS $\phi$ field is determined by non-linearly with $\rho_k$.    
\beq
\rho_{kn} = {\nabla f \grad \phi(x,y,l) \over 4 \pi G} =
{\nabla^2 [\Phi -\phi(x,y,l)] \over 4 \pi G}
\eeq

E.g., a TeVeS point lens has a non-zero convergence 
due to the non-zero effective DM halo.
So important differences between lensing in TeVeS and in GR is
in the predicted metric or potential $\Phi$ for a given galaxy mass
distribution $\rho_k$.

In general, our non-linear Poisson equation can be solved by 
adapting the code of Ciotti, Nipoti, Pasquale (2006).  
In some cases, one can also take the potential-to-density, 
and start with a reasonable guess for the potential, 
and find the density by taking appropriate derivatives, e.g., the 
application of Angus et al. (2006a, 2006b) on the bullet cluster. 
In special cases, one can also solve the TeVeS Poisson equations
analytically.  This is the case for Kuzmin disks.

\subsection{$\kappa$ in a non-spherical model}

Here we illustrate how to model non-spherical 
lens galaxy in TeVeS, and point out some subtle difference with DM.
We present models with identical rotation curves, 
one in TeVeS and one in DM, demonstrate the subtle difference
in their lensing signal.  

To keep the calculations tractable,
we approximate the lens potential as that of an edge-on Kuzmin disk
with a density 
\beq
\rho(R,Z) \equiv {M b \delta(Z) \over 2\pi (R^2+b^2)^{3/2} }
\eeq
in cylindrical coordinates around the symmetry axis $(R,Z)$ of the 
Kuzmin disk of total mass $M$.
We work out the bending angle and the lens equation in the
sky directions x and y. 
Here the Newtonian gravity ${\bf g}$ is given by, 
\bey
-(g^N_R, g^N_Z) & = & {G M \over \left[R^2+(Z+b_\pm)^2\right]^{3/2} } (R, Z + b_\pm ) 
\eey
where $b_\pm = \pm b$ for positive or negative $Z$ respectively,
because the upper part of a Kuzmin disk is in
a spherical potential like that of a point-mass centered on $(x_0,y_0,z_0)$
below the disk.   The potential of the lower part of the Kuzmin disk 
is a mirror-image of the upper part.  
The Kuzmin scale-length $b=\sqrt{x_0^2+y_0^2+z_0^2}$ 
is smaller than the half-light radius $r_{half}$ by a factor 
${\sqrt{3}}$.

The nice thing about a Kuzmin disk is that the MONDian gravity is parallel
to the Newtonian gravity riigourously because 
the equal potential contours coincide with equal 
Newtonian gravity contours.  
Assuming the Bekenstein mu-function, or 
\beq
\mu_s = {g^s \over a_0}, \qquad \mu_s {\mathbf g}^s = {\mathbf g}^N,
\eeq
we get the TeVeS scalar field
\bey
-(g^s_R, g^s_Z) & = & {v_0^2 \over R^2+(Z+b_\pm)^2 } (R, Z + b_\pm ) 
\eey
where
\beq
v_0^2 \equiv \sqrt{G M a_0}.
\eeq
One can easily verify that
\beq
-\grad \left[ \left| {{\mathbf g}^s \over a_0} \right| {\mathbf g}^s\right] = 
{4\pi G \rho}, \qquad 
\eeq  

One can also replace the scalar field with a spherical DM halo 
potential $v_0^2/2 \ln(R^2+Z^2+b^2)$ with a gravity
\bey
-(g^{DM}_R, g^{DM}_Z) & = & {v_0^2 \over R^2+Z^2+b^2 } (R, Z)
\eey
such that they two descriptions give identical gravity $g_R$ 
on the equtorial $Z=0$ plane where rotation curves are measured.  
So the circular velocity curve is 
\beq
V_{\rm cir}^2 =R g_R(R,0) =
= {G M R^2 \over \left[R^2+ b^2\right]^{3/2} } +  {v_0^2 R^2 \over R^2+b^2 } 
\eeq  
which is rising as solid-body at centre and asymptotically flat velocity
$v_0$ at large radii.  Far away from the disk plane, 
TeVeS resembles a spherical DM halo very well (Read \& Moore 2005).

Nonetheless there is a subtle difference.
While the DM force is centrally-pointing,
the scalar force is pointed more to a point below/above the disk.  
At large radii $R$, and small $|Z|$,  we have
\beq
|g^s_Z| - |g^{DM}_Z| 
\sim {b \sqrt{G M a_0} \over R^2+b^2},~{\rm |Z| \ll b < R} 
\eeq
so the TeVeS scalar field provides *stronger vertical force* than the 
spherical DM close to the disk plane.

\subsection{Lensing by Kuzmin Disk}

Assuming Bekenstein's $\mu$ function, we can get the lens equations,
\begin{equation}
x-x_s = {D_l D_{ls} \over D_s} \alpha_x =(x-x_0)k_1+(x+x_0)k_2,
\end{equation} 
\begin{equation}
y-y_s = {D_l D_{ls} \over D_s} \alpha_y =(y-y_0)k_1+(y+y_0)k_2,
\end{equation} 
where $(x_s,y_s)$ are the source positions (projected on the lens plane), 
$k_1$ and $k_2$ are dimensionless 
functions of the baryon mass, distances and inclinations.  

Consider the simplest case 
we have a face-on Kuzmin disk with the parameters $x_0=y_0=0$, $z_0=b$.
For this axisymmetric lens, deflection points to the lens origin
and is a function of $R$ only.   One can predict convergence and shear
by taking appropriate derivatives.  
The critical (Einstein) ring can be obtained by setting $x_s=y_s=0$ so
that the Jacobian $A^{-1}={dx_sdy_s \over dx dy}=0$.  

Let 
\beq
R_M \equiv \sqrt{{4 GM \over c^2} {D_l D_{ls} \over D_s}},
\qquad D_0 \equiv {c^2 \over a_0} \approx {6 c \over H_0} 
\eeq
be the Einstein ring size in GR and a characteristic MOND distance
respectively,
then TeVeS predicted critical radius $R$ satisfies
\beq
1 = {D_l D_{ls} \over D_s} {\alpha(R) \over R} = 2 (k_3+k_4), 
\eeq 
where 
\bey
2 k_3 &\equiv& {R_M^2 \over R^2} {\chi^2 \over 1+\sqrt{1-\chi^2}}\\
2 k_4 &\equiv& (2\arcsin{\chi} )
{R_M \over R} {D_l D_{ls} \over D_s D_0}\\
2 k_5 &\equiv& (\pi \chi) {R_M \over R} {D_l D_{ls} \over D_s D_0}\\
\chi &\equiv& {R \over \sqrt{R^2+b^2}}, 
\eey
where we have used $k_1=k_2=k_3+k_4$.  

The corresponding face-on Kuzmin disk plus spherical DM model predicts
\beq
1 = {D_l D_{ls} \over D_s} {\alpha(R) \over R} = 2 (k_3+k_5),
\eeq
Hence MOND can create an illusion of 
an isothermal halo lens with a terminal velocity $v_0^2=\sqrt{GMa_0}$
through the terms with $k_4$ and $k_5$.

Note $k_5 \ge k_4$.  This means 
we have a slightly *bigger* critical radius for 
the case with DM halo vs. the case with TeVeS.
Likewise TeVeS would predict slightly different time delay 
between images than DM halos for a fixed $H_0$.  
These are somewhat surprising since
the two models have {\it identical rotation curves}.  This suggests
that a combination of lensing and kinematics will be able to
differentiate DM and TeVeS.

\subsection{MONDian effects on critical rings}

To understand the correction due to MOND, let's consider 
lensing by a spherical point lens, which can be obtained 
by letting $b=0$, hence $\chi=1$, 
\beq
2k_3 ={R_M^2 \over R^2}, \qquad 2k_4=2k_5={\pi R_M \over R} 
{D_l D_{ls} \over D_0 D_s}
\eeq
It is helpful to estimate rescaled lens-source distance in TeVeS
 \beq
 {D_l D_{ls} \over D_0 D_s} \sim {D_l D_{ls} \over 25 {\rm Gpc} D_s} 
  \ll 0.1
 \eeq
for all lenses and sources. Hence $2k_4 \le 2k_5 \sim {D_lD_{ls} \over D_0D_s} \ll 2k_3 \sim 1$ near the 
critical ring $R\sim R_M$, 
so the MONDian effect is never very important near critical line.  
This means critical rings (or strong lensing) 
can only occur in line of sight which pass through regimes
of strong gravity  $R \sim R_M \ll \sqrt{G M \over a_0}$ 
where *the MONDian effect is small*.
 
\section{Lensing Tests of TeVeS}

For lenses with almost co-linear double images in the CASTLES survey, 
Zhao, Bacon, Taylor, Horne (2006) conducted a detailed fit using 
spherical point or Hernquist profile lenses.  Cares have been taken 
in including the K-correction, the luminosity evolution with redshift, and
the possibility of significant gas and extinction from dust.
They applied two methods, using the image positions only, and using
the image amplifications.  
They found that
the mass-to-$M_*$ ratios calculated using the two independent methods
closely agree, and all but two of the lenses are found to
have $M/M_*$ between 0.5 and 2. This
shows that TeVeS is a sensible theory for doing gravitational lensing.

However, the authors caution that there are several lenses in galaxy
clusters which require extremely high M/L.  Clearly more detailed
models are needed, including flattening and the cluster environment.
As a first attempt in this direction, Angus et al. (2006) found that
the lensing peaks of the Bullet Cluster could be explained by adding
neutrinos in a TeVeS-like modified gravity; the phase space density of
neutrinos at the lensing peaks requires 2eV mass to in order not to
violate exclusion principle for fermions.  In general gravitational
lensing can be used as a useful approach to distinguish between
theories of gravity, and to probe the functional form of the
modification function ${\mu}$.

\section{Other Sanity Checks of TeVeS}
 
Sanity checks from small to large scale have also been done 
in recent papers.  TeVeS is found to be

\begin{itemize}
\item OK with solar system (Bekenstein \& Maguijo 2006)  
\item OK with Milky Way and Bulge Microlensing (no cusp problem, Famaey \& Binney 2006)
\item Excellent description of spiral rotation curves (McGaugh 2005, Famaey et al. 2006)
\item OK with elliptical galaxies lenses (Zhao, Bacon, Taylor, Horne 2006)
\item OK with galaxy clusters if with neutrinos (Angus, Shan, Zhao, Famaey, 2006),
\item TeVeS universe can accelerate (Zhao 2006, astro-ph/0610056)
\item Structures and CMB can form from linear perturbations (Dodelson \& Liguori 2006).  
\end{itemize}

TeVeS is by no means a firmed established paradigm since
many comparisons of the theory with observations are still unknown, 
but in the process of understanding and falsifying TeVeS, 
we learn to design clever dark matter
models and appreciate better the robustness of GR.
As it stands, 
\begin{itemize}
\item TeVeS is not grossly inconsistent with observations of lensing 
apart from a few outliers associated with galaxy clusters where massive 
neutrinos would contribute to the deflection of the light,
\item CMB anisotropy are predictable (Skordis et al. 2005), 
\item structure formation in non-linear potential can in principle be followed by N-body codes (Ciotti et al. 2006).  
\end{itemize}
 
\section*{Acknowledgements}
I thank Benoit Famaey, Pedro Ferriera, Tom Zlonik, 
Constantinous Skordis, David Mota, Luca Ciotti, Carlo Nipoti
Huanyuan Shan, Daming Chen for discussions 
on cosmology, dynamics and lensing.
The material of this presentation draws from following papers:



\begin{thebibliography}{}
\bibitem{} 
Angus G., Shan H, Zhao H, Famaey B. 2006, ApJ Letter, in press (astro-ph/0609125)
\bibitem{} 
Bekenstein J. 2004, Phys. Review D., 70, 3509
\bibitem{} 
Bekenstein J. \& Maguijo J. 2006, Phys. Review D., 
\bibitem{} 
Bradac M. et al. 2006, ApJ, in press (astro-ph/0608408)
\bibitem{} 
Ciotti L. , Londrillo P., Nipoti C. 2006, ApJ, 640, 741
\bibitem{}
Clowe D., et al. 2006, ApJ, 648, L109
\bibitem{}
Famaey B., Binney J. 2005, 363, 603
\bibitem{}
Famaey B., Gianfranco G., Bruneton J.P., Zhao H. 2006, Phy.Rev.D, (astro-ph/0611227)
\bibitem{} 
Milgrom M. 1983, ApJ, 270, 365
\bibitem{} 
McGaugh S. 2005, PRL
\bibitem{}
Read J, Moore B. 2005, MNRAS, 361, 971
\bibitem{}
Zhao H., Bacon D., Taylor A., Horne K. 2006, MNRAS, 368, 171
\bibitem{}
Zhao H., review for ``Quantum to cosmology: fundamental Physics in Space'',  Washinton, astro-ph/0610056
\bibitem{}
Zhao H.  Famaey B. 2006, ApJ Letter, 638, L9
\bibitem{}
Zlonik, Ferreira, Starkman 2006, Physical Review Letter, (astro-ph/0607441).
\bibitem{}
Zlonik, Ferreira, Starkman 2006, Physical Review D., 74, 0404037
\end{thebibliography}
\end{document}